%% file: main.tex
\documentclass{llncs}

\usepackage{times,amsmath,amsfonts,epsfig}
\usepackage{color}
\usepackage{tikz}
\usetikzlibrary{shapes,decorations,arrows,snakes,positioning,calc,shadows}
\usepackage{graphicx}
\usepackage{etoolbox}
\usepackage{cite}
\usepackage{enumerate}
\usepackage{url}
\usepackage{enumitem}
\usepackage[T1]{fontenc}
\usepackage[strings]{underscore}

\input{macros}

\makeatletter
\renewcommand*{\@fnsymbol}[1]{\ensuremath{\ifcase#1\or *\or \dagger\or \ddagger\or
   \mathsection\or \mathparagraph\or \|\or **\or \dagger\dagger
   \or \ddagger\ddagger \else\@ctrerr\fi}}
\makeatother

\begin{document}

\title{Constructing Data Graphs for Keyword
Search\thanks{This work was supported by the Israel Science Foundation (Grant No.~1632/12).} \thanks{This paper is the full version of~\cite{dexawGS16}.
The final publication is available at Springer via http://dx.doi.org/10.1007/978-3-319-44406-2\_33.}}

\author{Konstantin Golenberg\inst{1} \and Yehoshua Sagiv\inst{2}}
\institute{The Hebrew University, Jerusalem 91094, Israel,\\
\email{konstg01@cs.huji.ac.il}
\and
The Hebrew University, Jerusalem 91094, Israel,\\
\email{sagiv@cs.huji.ac.il}
}

\maketitle              

\begin{abstract} 
  A data graph is a convenient paradigm for supporting keyword search
  that takes into account available semantic structure and not just
  textual relevance.  However, the problem of constructing data graphs
  that facilitate both efficiency and effectiveness of the underlying
  system has hardly been addressed. A conceptual model for this task
  is proposed. Principles for constructing good data graphs are explained.
  Transformations for generating data graphs from RDB
  and XML are developed. The results obtained from these
  transformations are analyzed. It is shown that XML is a better
  starting point for getting a good data graph.
\end{abstract}


\input{intro}
\input{prelim}
\input{why_directed}
\input{advantages}

\input{constructing}
\input{comparing}
\input{conclusion}

{\small
\bibliographystyle{splncs}
\bibliography{main}
}

\end{document}

%% file: macros.tex
\definecolor{type}{RGB}{17,138,145}
\definecolor{attr}{RGB}{22,49,255}
\definecolor{val}{RGB}{240,56,240}

\newcommand{\xmltype}[1]{\texttt{\textcolor{type}{#1}}}
\newcommand{\xmlattr}[1]{\texttt{\textcolor{attr}{#1}}}
\newcommand{\xmlval}[1]{\texttt{\textcolor{val}{#1}}}
\newcommand{\xmlav}[2]{\xmlattr{#1=}\xmlval{"#2"}}
\newcommand{\xmlelem}[2]{\xmltype{<#1>}\texttt{#2}\xmltype{</#1>}}
\newcommand{\tab}{~~~~}
\newcommand{\ltab}{~~~~~~~~~~~~~~~~~~}
\newcommand{\dtdelem}[1]{\xmltype{<!ELEMENT} \xmlattr{#1} \xmlval{(\#PCDATA)}\xmltype{>}}
\newcommand{\stl}{\bf\footnotesize}

\newcommand{\theader}[1]{\multicolumn{1}{c|}{\textbf{#1}}}
\newcommand{\tkey}[1]{\multicolumn{1}{c|}{\textbf{\underline{\smash{#1}}}}}
\newcommand{\theaderl}[1]{\multicolumn{1}{c}{\textbf{#1}}}

\makeatletter
\newcommand{\labitem}[2]{%
\def\@itemlabel{\textbf{#1}}
\item
\def\@currentlabel{#1}\label{#2}}
\makeatother

%% file: intro.tex
\section{Introduction}
Considerable research has been done on effective algorithms
for keyword search over data graphs
(e.g.,~\cite{CW14,icdeBHNCS02,vldbKPCSDK05,sigmodGSBS03,sigmodHWYY07,sigmodGKS08,icdtGS16,edbtMS16}).
Usually, a data graph is obtained from RDB, XML or RDF by a rather
simplistic transformation.  In the case of RDB~\cite{icdeBHNCS02,
  icdeDYWQZL07, sigmodHWYY07}, tuples are nodes and foreign keys are
edges.  When the source is XML~\cite{sigmodGSBS03, icdeHPB03},
elements are nodes, and the edges reflect the document
hierarchy and IDREF(S) attributes.  

In many cases, the source data suffers from certain anomalies and some
papers (e.g.,~\cite{icdeHPB03,icdeKRSSW09}) take necessary steps to
fix those problems.  For example, when citations are represented by
XML elements, they should be converted to IDREF(S) attributes.  As
another example, instead of repeating the details of an author in each
one of her papers, there should be a single element representing all
the information about that author and all of her papers should
reference that element. These are examples of necessary
transformations on the source data. If they are not done, existing
algorithms for keyword search over data graphs will not be able to
generate meaningful answers.

Once a source data is ameliorated, it should be transformed into a graph.
The literature hardly discusses how it should be done.
In~\cite{icdeBHNCS02,vldbKPCSDK05}, the source is an RDB
and the naive approach mentioned earlier is used
(i.e.,~tuples are nodes and foreign keys are edges).
In~\cite{sigmodGSBS03,edbtXP08}, the source data is XML
and the simplistic transformation described at the beginning of this section
is applied.
In~\cite{pvldbDKS08,sigmodLOFWZ08,dblpBLCL09,ipmPL15,sigmodHWYY07},
they do not mention any details about the construction of data graphs.
The lack of a thoughtful discussion in any of those papers is rather surprising, because the
actual details of constructing a data graph have a profound effect on both
the efficiency and the quality of keyword search, regardless of the
specific algorithms and techniques that are used for generating answers and ranking them.

Construction of effective 
data graphs is not a simple task, since the
following considerations should be taken into account.  For
efficiency, a data graph should be as small as possible. It does not
matter much if nodes have large textual contents, but the number of
nodes and edges is an important factor.  However, lumping together
various entities into a single node is not a good strategy for
increasing efficiency, because answers to queries would lose their
coherence.

The structure of a data graph should reflect succinctly the semantics
of the data, or else answers (which are subtrees) would tend to be
large, implying that finding them would take longer and 
grasping their meaning quickly would not be easy.

An effective engine for keyword search over data graphs must also use
information-retrieval techniques. Those tend to perform better on large chunks
of text, which is another reason against nodes with little content.

In this paper, we address the problem of how to construct data graphs
in light of the above considerations. In
Section~\ref{sec:constr-data-graphs}, we develop transformations for
constructing data graphs from RDB and XML. 
In Section~\ref{sec:comparing}, we show that the format of the source data
(i.e.,~RDB or XML) has a significant impact on the quality of
the generated data graph. Moreover, XML is a better starting point
than RDB.  This is somewhat surprising given the extensive research
that was done on designing relational database schemes. 

As a conceptual guideline for constructing a good data graph,
we use the OCP model~\cite{sigmodAGKS10}, which was developed for supporting
a graphical display of answers so that their
meaning is easily understood. In Section~\ref{sec:advantages},
we explain why the OCP model is also useful as a general-purpose basis
for constructing data graphs in a way that takes into account all
the issues mentioned earlier.

In summary, our contributions are as follows. First, we enunciate the
principles that should guide the construction of data graphs. Second,
we develop transformations for doing so when the source data is RDB or
XML. These transformations are more elaborate than the simplistic
approach that is usually applied.  Third, we show how the format of
the source data impacts the quality of the generated graphs. Moreover,
we explain why XML is a better starting point than RDB.

Our contributions are valid independently of a wide range of issues
that are not addressed in this paper, such as the algorithm for
generating answers and the method for ranking them. 
We only assume
that an answer is a non-redundant subtree that includes all the keywords
of the query. However, our results still hold even if answers are
subgraphs, as sometimes done.

A presentation that gives motivation for the work of this paper is given in~\cite{dexawGS16}.

%% file: prelim.tex
\section{Preliminaries}\label{sec:prelim}
\subsection{The OCP Model}\label{sec:ocp}
The \emph{object-connector-property} (OCP) model for data graphs was developed
in~\cite{sigmodAGKS10} to facilitate an effective GUI for presenting subtrees.
(As explained in the next section, those subtrees are answers to keyword search
over data graphs.)  In the OCP model, objects are entities and
connectors are relationships.  We distinguish between two kinds of
connectors: \emph{explicit} and \emph{implicit}. Objects and explicit
connectors can have any number of properties.
Two special properties are \emph{type} and \emph{name}.

\input{fig_ocp2}

Parts~(a) and~(b) of Figure~\ref{fig:ocp} show an object and a
snippet of a data graph, respectively.  An object is depicted as
a rectangle with straight corners. The top line of the rectangle
shows the name and type of the object.  The former appears first
(e.g.,~\texttt{Ukraine}) and the latter is inside parentheses
(e.g.,~\texttt{country}).  The other properties appear as pairs consisting of
the property's name and value, as shown in
Figure~\ref{fig:ocp}(a). Observe that properties can be nested; for
example, the property \texttt{percentage} is nested inside \texttt{ethnicgroup}. Nesting
is indicated in the figure by indentation. 

An implicit connector is shown as a directed edge between two objects.
Its meaning should be clear from the context.  In
Figure~\ref{fig:ocp}(b), the implicit connector from \texttt{Ukraine}
to \texttt{Odeska} means that the latter is a province in the former. 

An explicit connector is depicted as 
a rectangle with rounded corners. It has at most one incoming edge from an
object and any positive number of outgoing edges to some objects.  An
explicit connector has a type, but no name, and may also possess other
properties. Figure~\ref{fig:ocp}(b) shows an explicit connector of type border
from \texttt{Ukraine} to \texttt{Russia} that has the
property \texttt{length} whose value is \texttt{1576km}. 

\subsection{Answers to Keyword Search}\label{sec:keyword-search}
We consider keyword search over a directed data graph $G$.  (A data graph must
be directed, because relationships among entities are not always symmetric.)
A \emph{directed
  subtree} $t$ of $G$ has a unique node $r$, called the \emph{root},
such that there is exactly one directed path from $r$ to each node of
$t$.

A query $Q$ over a data graph $G$ is a set of keywords, namely,
$Q=\{k_1,\ldots,k_n\}$.  An \emph{answer} to $Q$ is a directed subtree $t$ of
$G$ that contains all the keywords of $Q$ and is nonredundant, in the
sense that no proper subtree of $t$ also contains all of them.

\input{fig_answers}

For example, consider Figure~\ref{fig:mondial_opp}, which shows a
snippet of the data graph created from the XML version of the Mondial
dataset,\footnote{\url{http://www.dbis.informatik.uni-goettingen.de/Mondial/}}
according to the transformation of Section~\ref{sec:xml}.
To save space, only the name (but not the type) of each object is shown.
The dashed edges should be ignored for the moment.  The subtree in
Figure~\ref{fig:answers}(a) is an answer to the query
$\{\texttt{Dnepr, Russia, Ukraine}\}$. There are additional answers to
this query, but all of them have more than three nodes and at least
one explicit connector.

For the query $\{\texttt{Dnepr, Don}\}$, there is no answer (with only
solid edges) saying that \texttt{Dnepr} and \texttt{Don} are rivers in \texttt{Russia}, although
the data graph stores this fact. The reason is that the connectors (in
the data graph of Figure~\ref{fig:mondial_opp}) have a symmetric
semantics, but the solid edges representing them are in only one
direction.  The only exception is the connector \texttt{border}, which
is already built into the graph in both directions (between
\texttt{Russia} and \texttt{Ukraine}).
In order not to miss answers, we add \emph{opposite edges} when
symmetric connectors do not already exist in both directions.  Those
are shown as dashed arrows. Now, there are quite a few answers to the
query $\{\texttt{Dnepr, Don}\}$ and Figure~\ref{fig:answers}(b)--(d)
shows three of them.  The first two of those say that \texttt{Dnepr}
and \texttt{Don} are rivers in \texttt{Russia}. These two answers have
the same meaning, because the relationship between a river and a
country is represented twice: by an implicit connector and by the
explicit connector \texttt{located}. The answer in
Figure~\ref{fig:answers}(d) has a different meaning, namely,
\texttt{Dnepr} and \texttt{Don} are rivers in \texttt{Ukraine} and
\texttt{Russia}, respectively, and there is a \texttt{border} between
these two countries.

To generate relevant answers early on, weights are assigned to the
nodes and edges of a data graph.  Existing algorithms
(e.g.,~\cite{icdeBHNCS02,sigmodGKS08,pvldbGKS11,icdtGS16} enumerate
answers in an order that is likely to be correlated with the desired
one.  Developing an effective weighting scheme is highly important,
but beyond the scope of this paper.

%% file: fig_ocp2.tex
\begin{figure}[t]
\centering
\begin{minipage}{0.35\textwidth}
\begin{tikzpicture}[->,>=stealth',shorten >=0pt,auto,node distance=10pt]
\tikzstyle{node}=[rectangle,draw,minimum size=20pt]
\tikzstyle{connector}=[rectangle,rounded corners, draw,minimum size=20pt]

	\node[node, align=left] (ukr1) at (0,0) {	  
		\textrm{Ukraine} {\scriptsize (country)} \\
		~~~~area : $603,500 \textrm{km}^2$ \\
		~~~~government : republic\\
		~~~~ethnicgroups : Ukrainian\\ 
		~~~~~~~~percentage : $73$	  
	};
	\node [left = of ukr1]{(a)};

	\node[node, align=left, below = 1.1 of ukr1] (ukr) {\textrm{Ukraine} {\scriptsize (country)}};
	\node[node, align=left, below right=.3 and -1 of ukr] (prov) {\textrm{Odeska} {\scriptsize (province)}};
	\node[connector, align=left, below left=.3 and -1 of ukr] (border) {	  
		{\textrm{border}}\\
		~~~~length : $1576\textrm{km}$
	};
	\node[node, align=left, below = of border] (rus) {\textrm{Russia} {\scriptsize (country)}};
	\node [left = 1.05 of ukr]{(b)};
	
	\draw (ukr) edge  (border);
	\draw (border) edge (rus);
	\draw (ukr) edge (prov);

\end{tikzpicture}
\caption{\label{fig:ocp}An object and a tiny snippet of a data graph (not all properties are shown)}
\end{minipage}
\hfill
\begin{minipage}{0.60\textwidth}
\begin{tikzpicture}[->,>=stealth',shorten >=0pt,auto,node distance=2cm]
\tikzstyle{node}=[rectangle,draw,minimum size=20pt]
\tikzstyle{connector}=[rectangle,rounded corners, draw,minimum size=20pt]
  
  \node[node] (1) at (0,0) {Russia};
  \node[node] (6) [left =2 of 1] {Ukraine};
  \node[connector] (2) [below of = 1]{located};
  \node[node] (3) [above right = 0.5 of 2]{Don};
  \node[connector] (9) [above right = 0.7 of 6]{border};
  \node[connector] (10) [below right = 0.7 of 6]{border};
  \node[node] (5) [above  of=9]{Dnepr};
  \node[connector] (7) [below = 0.45 of 5]{located};

  \draw (2) [bend right = 10] edge (1);
  \draw (3) [bend left = 20] edge (2);
  \draw (3) edge [bend left = 10](1);
  \draw (5) edge [bend right] (7);
  \draw (6) edge [bend left] (9);
  \draw (9) edge [bend left] (1);
  \draw (1) edge [bend left] (10);
  \draw (10) edge [bend left] (6);
  \draw (7) [bend left = 50] edge (1);
  \draw (7) [bend right = 50] edge (6);
  \draw (5) edge [bend left = 80] (1);
  \draw (5) edge [bend right= 80] (6);
  
  \draw (6) edge [dashed, bend left = 90] (5);
  \draw (1) edge [dashed, bend right= 90] (5);
  
  \draw (1) [dashed, bend right = 30] edge (7);
  \draw (6) [dashed, bend left = 30] edge (7);
  
  \draw (7) [dashed, bend right = 20] edge (5);
  
  \draw (1) [dashed, bend right = 10] edge (2);
  \draw (2) [dashed, bend left = 20] edge (3);
  \draw (1) [dashed, bend left = 10] edge (3);

\end{tikzpicture}
\caption{\label{fig:mondial_opp}A tiny portion of Mondial}
\end{minipage}
\end{figure}

%% file: fig_answers.tex
\begin{figure}[t]
\centering
\begin{tikzpicture}[->,>=stealth',shorten >=0pt,auto,node distance=1.5cm]
\tikzstyle{node}=[rectangle,draw,minimum size=20pt]
\tikzstyle{connector}=[rectangle,rounded corners, draw,minimum size=20pt]
  \node[node] (d1) at (0,0) {Dnepr};
  \node[node] (r1) [below left of = d1] {Russia};
  \node[node] (u1) [below right of = d1] {Ukraine};
  \draw (d1) edge (r1);
  \draw (d1) edge (u1);
  \node [below = 1.2 of d1] {(a)};
  
  \node[node] (r2) at (4,0) {Russia};
  \node[node] (dn2) [below left of = r2] {Denpr};
  \node[node] (do2) [below right of = r2] {Don};
  \draw (r2) [dashed] edge (dn2);
  \draw (r2) [dashed] edge (do2);
  \node [below = 1.2 of r2] {(b)};
  
  \node[node] (r3) at (0,-3.2) {Russia};
  \node[node] (dn3) [below left of = r3] {Denpr};
  \node[connector] (ldo3) [below right of = r3] {located};
  \node[node] (do3) [right of = ldo3] {Don};
  \node [below = 1.2 of r3] {(c)};
    
  \draw (r3) [dashed] edge (dn3);
  \draw (r3) [dashed] edge (ldo3);
  \draw (ldo3) [dashed] edge (do3);
  
  \node[node] (u4) at (8,0) {Ukraine};
  \node[node] (dn4) [below left of = u4] {Dnepr};
  \node[connector] (b4) [below right of = u4] {border};
  \node[node] (r4) [below of = b4] {Russia};
  \node[node] (do4) [below of = r4] {Don};

  \node [below = 4 of u4] {(d)};
    
  \draw (u4) [dashed] edge (dn4);
  \draw (u4) edge (b4);
  \draw (b4) edge (r4);
  \draw (r4) [dashed] edge (do4);

\end{tikzpicture}
\caption{\label{fig:answers}Answers to queries}
\end{figure}
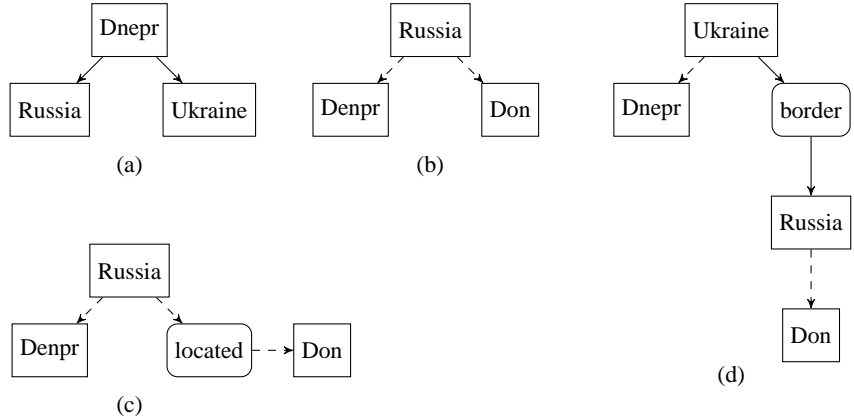

%% file: why_directed.tex
\input{fig_asym_conn}
\subsection{Why Data Graphs are Directed}\label{sec:why}
Data graphs must be directed because some relationships are asymmetric.
For example, to represent citations among papers, we need two different
types of connectors, as shown in Figure~\ref{fig:asym_conn}.
In contrast, one connector type is sufficient for representing borders.

\input{fig_border}
When a relationship is symmetric, it is redundant to use a different
connector node for each direction, which is the case with \texttt{border}
in Figure~\ref{fig:mondial_opp}. It is better to represent a border
between two countries as in Figure~\ref{fig:border}.

Over the data graph of Figure~\ref{fig:border},
the following are exactly two answers to the query $\{\texttt{Russia, Ukraine}\}$.
\begin{itemize}[noitemsep,nolistsep]
\item
$\texttt{Ukraine} \rightarrow \texttt{border} \rightarrow \texttt{Russia}$
\item
$\texttt{Russia} \dashrightarrow \texttt{border} \dashrightarrow \texttt{Ukraine}$
\end{itemize}
As directed subtrees, these answers are distinct. However, they carry
the same information.  
Hence, we eliminate duplicates (similarly to~\cite{icdeBHNCS02}) by
treating an answer as a set of undirected edges.  That is, two answers
are the same if they have the same set of undirected edges.  Equality
of undirected edges is determined as follows.  Each node has a unique
id (which is internal to the system). Thus, two edges are identical
if they are the same unordered pair of id's.

Over the data graph of Figure~\ref{fig:mondial_opp}, however, there
are two distinct \texttt{border} nodes between \texttt{Ukraine} and
\texttt{Russia}.  Hence, the following two answers
\begin{itemize}[noitemsep,nolistsep]
\item
$\texttt{Ukraine} \rightarrow \texttt{border} \rightarrow \texttt{Russia}$
\item
$\texttt{Russia} \rightarrow \texttt{border} \rightarrow \texttt{Ukraine}$
\end{itemize}
are distinct even when viewed as undirected subtrees.  To eliminate
duplicates also in this case, we need to consider two connector nodes
as equal if they have the same type, rather than the same id.

Even when a connector type is asymmetric, it is redundant to present
both directions. For example, given the data graph of
Figure~\ref{fig:asym_conn}, the following two subtrees carry the same
information, in spite of having nodes of different types.
\begin{itemize}[noitemsep,nolistsep]
\item
$\texttt{Paper A} \rightarrow \texttt{cite} \rightarrow \texttt{Paper B}$
\item
$\texttt{Paper B} \rightarrow \texttt{cited\_by} \rightarrow \texttt{Paper A}$
\end{itemize}
To eliminate one of these two as a duplicate, we need to treat two 
connector nodes as equal if one has the inverse type of the other.

%% file: fig_asym_conn.tex
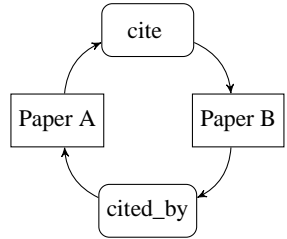
\begin{figure}[t]
\centering
\begin{tikzpicture}[->,>=stealth',shorten >=0pt,auto,node distance=1.7cm]
\tikzstyle{node}=[rectangle,draw,minimum size=20pt]
\tikzstyle{connector}=[rectangle,rounded corners, draw,minimum size=20pt]
  
  \node[node] (1) at (0,0) {Paper A};
  \node[connector] (2) [below right of = 1] {cited\_by};
  \node[connector] (3) [above right of = 1] {~~~cite~~~~~};
  \node[node] (4) [below right of = 3] {Paper B};

  \draw (1) edge [bend left](3);
  \draw (2) edge [bend left](1);
  \draw (3) edge [bend left](4);
  \draw (4) edge [bend left](2);
\end{tikzpicture}
\caption{\label{fig:asym_conn}An asymmetric connector and its inverse}
\end{figure}

%% file: fig_border.tex
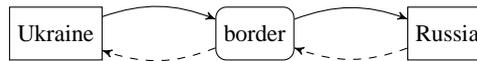
\begin{figure}[t]
\centering
\begin{tikzpicture}[->,>=stealth',shorten >=0pt,auto,node distance=1.5cm]
\tikzstyle{node}=[rectangle,draw,minimum size=20pt]
\tikzstyle{connector}=[rectangle,rounded corners, draw,minimum size=20pt]

  \node[node] (u) at (0,0) {Ukraine};
  \node[connector] (b) [right = of u] {border};
  \node[node] (r) [right = of b] {Russia};

  \draw (u) [bend left = 20] edge (b);
  \draw (b) [dashed, bend left = 20] edge (u);
  \draw (r) [dashed, bend left = 20] edge (b);
  \draw (b) [bend left = 20] edge (r);
\end{tikzpicture}
\caption{\label{fig:border}A single connector node for border}
\end{figure}

%% file: advantages.tex
\section{Advantages of the OCP Model}\label{sec:advantages}
In this section, we discuss some of the advantages of the OCP model.
In a naive approach of building a data graph, there is only one type of nodes
(i.e.,~no distinction between objects and connectors). Moreover, 
sometimes there is even a separate node for each property.
This approach suffers from three drawbacks.
First, from the implementation's point of view, this is inefficient in
both time and space. That is, even if there is not much data, the
number of nodes and edges is likely to be large. As a result,
searching a data graph for answers would take longer (than the
alternative described later in this section). In addition, if all the
processing is done in main memory, the size of the data graph is more likely
to become a limiting factor.

The second drawback of the naive approach is from the user's point of
view.  A meaningful answer is likely to have quite a few nodes; hence,
displaying it graphically in an easily understood manner is rather
hard.  Another problem is the following.  The definition of an answer
is intended to avoid redundant parts in order to cut down the search
space. However, sometimes an answer must be augmented to make it clear
to the user.  For example, an answer cannot consist of just some
property that contains the keywords of the query, without showing the
context.

The third drawback pertains to ranking, which must take into account
textual relevance (as well as some other factors). In the naive
approach, many nodes have only a small amount of text,
making it hard to determine their relevance to a given query.

In comparison to the naive approach, the OCP model dictates \emph{fat}
nodes.  That is, an object or an explicit connector is represented by
a node that contains all of its properties.  Consequently, we get the following
advantages. First, a data
graph is not unduly large, which improves efficiency.  Second, relevance is
easier to determine, because all the text pertaining to an object or
an explicit connector is in the same node.  Third, the GUI
of~\cite{sigmodAGKS10} is effective, because it does not clutter the
screen with too many nodes or unnecessary stuff. In particular, the default
presentation of an answer 
is condensed and only shows:
types and names of objects; types of explicit connectors; and properties that match some keywords of the query.
The user can optionally choose an expanded view in order to see all
the properties of the displayed nodes, when additional information
about the answer is needed.  Since all the properties are stored in
the nodes that are already shown, this can be done without any delay.
Furthermore, the GUI of~\cite{sigmodAGKS10} visualizes 
the conceptual distinction 
between objects and connectors, which makes it much easier to quickly grasp the meaning of an
answer.

%% file: constructing.tex
\section{Constructing Data Graphs\label{sec:construct}}\label{sec:constr-data-graphs}
\input{rel}
\input{xml}

%% file: rel.tex
\subsection{Relational Databases to Data Graphs}\label{sec:rel}
The naive approach for transforming a relational database into a data
graph (e.g.,~\cite{icdeBHNCS02}) does not distinguish between objects
and connectors.  For each tuple $t$, a node $v_t$ is created, such
that the relation name of $t$ is the type of $v_t$.  An edge from
$v_{t_1}$ to $v_{t_2}$ is introduced if tuple $t_1$ has a foreign key
that refers to $t_2$. Finally, opposite edges are also added.  In this
section, we describe more elaborate rules that create a data graph
with fat nodes and a clear distinction between objects and connectors.

As a matter of terminology, when we say ``foreign key $F$,'' we mean
that the foreign key consists of the set of attributes $F$.
A foreign key $F$ is transformed to a connector. Whether that
connector is implicit or explicit depends on the names of the
attributes comprising $F$. For example, suppose that there is a
relation named \texttt{Student}. If the attribute \texttt{student} is
a foreign key that points to that relation, then it can be transformed
to an implicit connector. However, if the attribute \texttt{grader}
points to the relation \texttt{Student}, it means that the foreign key
corresponds to an entity that has a special role and is not just an
ordinary student. In this case, the translation should create an explicit
connector of type \texttt{grader}.

The above example serves as a motivation for the following definition.
Suppose that $F$ is a foreign key that refers to a relation $R$.  Let
$P$ be the set consisting of the attributes of the primary key of $R$
and $R$ itself (i.e.,~the name of the relation). We say that the
foreign key $F$ is \emph{insignificantly named} if $F \subseteq P$;
otherwise, it is \emph{significantly named}. For example, let $F =
\{\texttt{student}\}$ be a foreign key that refers to the relation
\texttt{Student} that has the primary key \texttt{id}; hence, $P =
\{\texttt{Student,id}\}$. $F$ is insignificantly named, because $F
\subseteq P$. In practice, it is sufficient that $F$ is similar
(rather than strictly equal) to a subset of $P$. For example, if $F =
\{ \texttt{student\_id}\}$, then we still deem $F$ insignificantly
named.  (Due to a lack of space, we do not discuss how to test such
similarity automatically.) If $F = \{\texttt{grader}\}$, then $F
\not\subseteq P$ and hence $F$ is significantly named.

Given a relation $R$, we transform its tuples to objects and connectors
according to one of the following four cases.
\begin{enumerate}
\item\label{pwof} The primary key of $R$ does not include any foreign keys.
\item\label{pofp} The primary key $K$ of $R$ includes a single foreign
  key $F$ and either $K$ has some attributes in addition to those of
  $F$ or $F$ is significantly named.
\item\label{ptmf} The primary key $K$ of $R$ is a combination of at least
  two foreign keys and possibly some other attributes.
\item\label{pof} The relation $R$ has exactly one foreign key $F$,
  which is insignificantly named and is also the primary key.
\end{enumerate}

\input{fig_rdb_example2}
\input{fig_rdb_datagraph2}

In Case~\ref{pwof}, \ref{pofp} and~\ref{ptmf}, a tuple of $R$ is
an entity, a weak entity and a relationship, respectively.
In these cases, we do the following.
Each tuple $t$ of $R$ is transformed to a node $v_t$.
In the first two cases, $v_t$ is an object.
In the third case, $v_t$ is either an explicit connector or an object,
according to the following rule.
If all the foreign keys of $R$ are insignificantly named,
then $v_t$ is an explicit connector; otherwise, we make $v_t$ an object (to avoid
the creation of two explicit connectors that are
adjacent).\footnote{Even when all the foreign keys of $R$ are
  insignificantly named, $v_t$ has to be an object if some other
  relation references tuples of $R$. However, since $R$ represents a set of
  relationships (rather than entities), this possibility is unlikely to occur.}

The type of $v_t$ is $R$ (i.e.,~the name of the relation).  The
properties of $v_t$ are all the attributes of $t$ that do not belong
to foreign keys.  If $v_t$ is an object, its name is chosen to be the
value of an appropriate property (e.g.,~title, name, etc.). In
particular, we prefer a meaningful name over some meaningless id, even
if the former does not uniquely identify the object.

In addition, for each foreign key $F$ of $t$, we do the following. Let
$v_{t[F]}$ be the object corresponding to the tuple referenced by
$t[F]$ (i.e.,~the value of $t$ for $F$).  If $F$ is
insignificantly named, we add a directed edge $e$ from $v_t$ to
$v_{t[F]}$ (note that $e$ is an implicit connector if $v_t$ is an object).  Otherwise (i.e.,~$F$ is significantly named), we create
an explicit connector $c^F_t$ of type $F$ and add the directed edges
$(v_t,c^F_t)$ and $(c^F_t,v_{t[F]})$.

In Case~\ref{pof}, the relation $R$ is an \emph{auxiliary table} that
provides additional information about the entities referenced by
$F$. We transform a tuple $t$ of $R$ to a nested property of the
object $v_{t[F]}$.  The top-level property is $R$ (i.e.,~the name of
the relation) and it nests all the attributes of $t$ that do not
belong to $F$. 

As an example, Figure~\ref{fig:rdb} shows a snippet of the Mondial
relational database. In each relation, the attributes of the primary
key are underlined and arrows show foreign keys. We now explain
how to construct the data graph of Figure~\ref{fig:rdb_datagraph}.
In this section, the dotted (implicit and explicit) connectors
of Figure~\ref{fig:rdb_datagraph} should be ignored.

The are two relations, namely, \texttt{River} and \texttt{Country},
that satisfy the condition of Case~\ref{pwof}.  For each one of their
tuples, Figure~\ref{fig:rdb_datagraph} has an object. Note that
\texttt{France} is chosen to be the name of an object, although it is
not the value of the primary key.  The relation \texttt{Economy}
satisfies the condition of Case~\ref{pof}. Therefore, its only tuple
becomes the nested property \texttt{economy} of \texttt{France}.

The relation \texttt{Province} satisfies the condition of
Case~\ref{pofp}.  Hence, the object \texttt{Rh\^one Alpes} of type
\texttt{province} is in Figure~\ref{fig:rdb_datagraph}; its only other
property is \texttt{area}.  The foreign key of \texttt{Province} is
insignificantly named, so we add an implicit connector from
\texttt{Rh\^one Alpes} to \texttt{France} that is shown as a dash-dotted arrow.  
Note that \texttt{country} is not a property of the object
\texttt{Rh\^one Alpes}, because it belongs to a foreign key.

The relation \texttt{Confluence} of Figure~\ref{fig:rdb} satisfies the
condition of Case~\ref{ptmf}. Two out of the three foreign keys
included in its primary key (i.e.,~\texttt{river1} and
\texttt{river2}) are significantly named. Hence, the single tuple of
\texttt{Confluence} is an object; however, there is a lack of an
attribute that can serve as the name of that object.  For each of the
two significantly named foreign keys, we add an explicit connector,
which is depicted using dash-dotted shapes (i.e.,~two arrows and
a rectangle).  The third foreign key comprises two attributes
(\texttt{province} and \texttt{country}) and is insignificantly
named. So, we add an implicit connector (shown as a solid arrow) from
the object \texttt{confluence} to the object \texttt{Rh\^one Alpes}.

The above example shows that a constructed data graph could deviate
from the original OCP model (of Section~\ref{sec:ocp}) in the following
way. There is an object (of type \texttt{confluence}) without
a name. It could be argued that this object should really be a
connector. However, the result would be a data graph with adjacent
connectors, which makes it harder to quickly grasp the meaning of
answers having them.  Moreover, a confluence actually corresponds to a
real-world entity.  In the RDB of Figure~\ref{fig:rdb}, it is a weak
entity. So, we can create a name by concatenating the values of some
primary-key attributes (e.g.,~\texttt{Rh\^one} and \texttt{Sa\^one}).

The \emph{original} edges are those created by the above
transformation.  We also add opposite edges (i.e.,~in the reverse
direction), because the semantic of foreign keys is inherently undirected.

%% file: fig_rdb_example2.tex
\begin{figure}[t]
\center
\setlength\tabcolsep{4.5pt}
\begin{tikzpicture}[,>=stealth',shorten >=0pt,auto,node distance=0.7cm]
   \node (c) [inner sep=0pt] at (0,0) {
	   \begin{tabular}{ l | l | l }
	    \multicolumn{3}{c}{Country}  \\ \hline
	    \tkey{code} & \theader{name}  & \theaderl{population} \\ \hline
	    F & France & 58M   
	   	\end{tabular}
   };
   
   \node (eco) [inner sep=0pt, above = 0.3 of c] {
	   \begin{tabular}{ l | l | l }
	   \multicolumn{3}{c}{Economy} \\ \hline
	    \tkey{country} & \theader{gdp}  & \theaderl{inflation} \\ \hline
	     F & \$$37,728$ & $1.7$\%   
	   \end{tabular}
   };
   
   \node (p) [inner sep=0pt, left = of c] {
	   \begin{tabular}{ l | l | l }
	   \multicolumn{3}{c}{Province} \\ \hline
	    \tkey{name} & \tkey{country} & \theaderl{area} \\ \hline
	    Rh\^one Alpes & F & 43698
	   	\end{tabular}
   };

   \node (r) [inner sep=0pt, above = 0.2 of p] {
	   \begin{tabular}{ l | l }
	   \multicolumn{2}{c}{River} \\ \hline
	    \tkey{name} & \theaderl{length}  \\ \hline
	    Sa\^one & $473$km \\
	    Rh\^one & $813$km \\
	    
	   	\end{tabular}
   };

   \node (con) [inner sep=0pt] at(-3,3.3){
	   \begin{tabular}{ l | l | l | l | l | l}
	   \multicolumn{5}{c}{Confluence} \\ \hline
	    \tkey{river1} & \tkey{river2} & \tkey{province} & \tkey{country} & \theader{lng} & \theaderl{lat}  \\ \hline
	    Rh\^one & Sa\^one & Rh\^one Alpes & F & $45^\circ 43'$N & $4^\circ 49'$E
	   	\end{tabular}
   };
   
   \draw  (c.north west) rectangle (c.south east);
   \draw  (eco.north west) rectangle (eco.south east);
   \draw  (eco.north west) rectangle (eco.south east);
   \draw  (p.north west) rectangle (p.south east);
   \draw  (r.north west) rectangle (r.south east);
   \draw  (con.north west) rectangle (con.south east);
	
   \draw [->]
   		 ([xshift=-45pt]eco.south)
   		 -- ++(0,-0.2)
   		 -- ++(-.7,0)
   		 |- ([yshift=-10pt]c);
   \draw [->]
   		 ([xshift=10pt]p.south)
   		 -- ++(0,-0.2)
   		 -- ++(2.2,0)
   		 |- ([yshift=-17pt]c);
   \draw [decorate,decoration={brace,amplitude=5pt}] 
         ([xshift=40pt, yshift=-3pt]con.south) -- ([xshift=-50pt, yshift=-3pt]con.south) node[midway](b){};
   \draw [->](b)
         -- ++(0,-0.2)
         -- ++(0.75,0)
         |- ([yshift=-14pt]p);
   \draw [->]
   		 ([xshift=-105pt]con.south)
   		 -- ++(0,-0.2)
   		 |- ([yshift=-22pt]r);
   \draw [->]
   		 ([xshift=-72pt]con.south)
   		 -- ++(-0,-0.1)
   		 -- ++(-1,0)
   		 |- ([yshift=-6pt]r);
\end{tikzpicture}
\caption{A snippet of the Mondial RDB}\label{fig:rdb}
\end{figure}
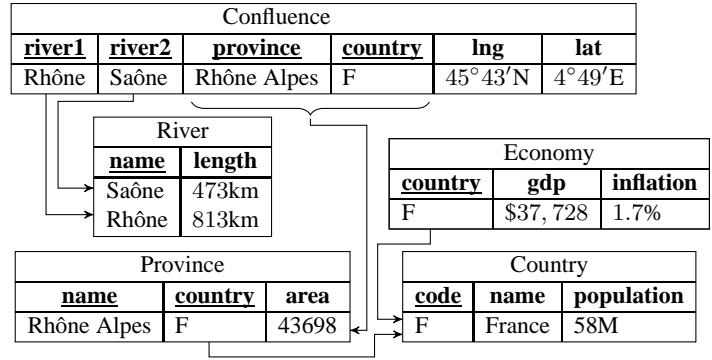

%% file: fig_rdb_datagraph2.tex
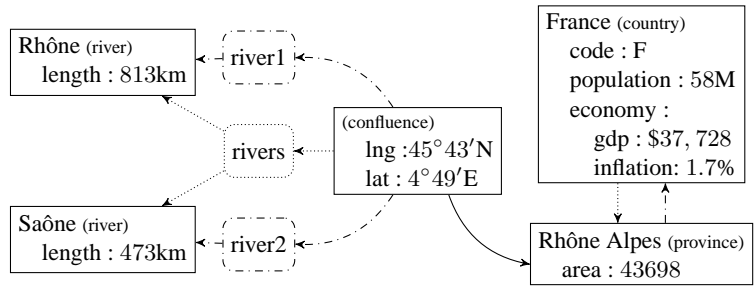
\begin{figure}[thb]
\centering
\begin{tikzpicture}[->,>=stealth',shorten >=0pt,auto,node distance=15pt]
\tikzstyle{node}=[rectangle,draw,minimum size=20pt]
\tikzstyle{connector}=[rectangle,rounded corners, draw,minimum size=20pt]

	\node[node, align=left] (con) at(0,0) {	  
		\textrm{\scriptsize (confluence)} \\
		\tab lng :$45^\circ 43'$N \\
		\tab lat : $4^\circ 49'$E
	};

	\node[connector, align=left, densely dotted] (rsc) [left = of con] {\textrm{rivers}};
	\node[connector, align=left, dashdotted] (rc1) [above = of rsc] {\textrm{river1}};
	\node[connector, align=left, dashdotted] (rc2) [below = of rsc] {\textrm{river2}};

	\node[node, align=left] (r1) [above left = of rsc] {	  
		\textrm{Rh\^one} {\scriptsize (river)} \\
		\tab length : $813$km 
	};

	\node[node, align=left] (r2) [below left = of rsc] {	  
		\textrm{Sa\^one} {\scriptsize (river)} \\
		\tab length : $473$km
	};

	\node[node, align=left] (p) [below right = of con] {	  
		\textrm{Rh\^one Alpes} {\scriptsize (province)} \\
		\tab area : $43698$ 
	};

	\node[node, align=left] (c) [above = of p] {	  
		\textrm{France} {\scriptsize (country)} \\
		\tab code : F \\
		\tab population : $58$M \\
	 	\tab economy : \\
	 	\tab\tab gdp : \$$37,728$ \\
	 	\tab\tab inflation: $1.7$\%
	};
	
	\draw (p) [dashdotted, transform canvas={xshift=1em}] edge  (c);
	\draw (c) [densely dotted, transform canvas={xshift=-1em}] edge  (p);
	\draw (con) [dashdotted, bend right] edge  (rc1);
	\draw (con) [dashdotted, bend left] edge  (rc2);
	\draw (rc1) [dashdotted] edge  (r1);
	\draw (rc2) [dashdotted] edge  (r2);

	\draw (con) [densely dotted] edge  (rsc);
	\draw (rsc) [densely dotted] edge  (r1);
	\draw (rsc) [densely dotted] edge  (r2);

	\draw (con) edge [bend right] (p);

\end{tikzpicture}
\caption{\label{fig:rdb_datagraph}A data graph constructed from the Mondial RDB}
\end{figure}

%% file: xml.tex
\subsection{From XML to Data Graphs}\label{sec:xml}

An XML document is a rooted hierarchy of \emph{elements}. Each element
can have any number of \emph{attributes}.  Three special types of
attributes are ID, IDREF and IDREFS. An attribute of the first type
has a value that uniquely identifies its element.  The last two types
serve as references to other elements. For an attribute defined (in
the DTD) as IDREF, the value is a single ID (of the referenced
element); and if an attribute is defined as IDREFS, its value is a set
of IDs.  In our terminology, a \emph{reference} attribute is one
defined as either IDREF or IDREFS. An attribute is \emph{plain} if it
is neither ID, IDREF nor IDREFS.

In XML lingo, an element has a \emph{name} that appears in its tag
(e.g.,~$<$city$>$). To avoid confusion, we call it the \emph{type} of the
element, because it corresponds to the notion of a type in the OCP model 

In this section, we describe how to transform an XML document to a
data graph.  We assume that the document has a DTD and use it in the
transformation.  As we shall see, the DTD provides information that is
essential to constructing the data graph. Conceivably, this information
can also be gleaned from the document itself. However, if the document
does not conform to a reasonable DTD, the resulting data graph
(similarly to the document itself) is likely to be poorly designed.
By only assuming that there is a DTD (as opposed to an XML schema),
we make our transformation much more applicable to real-world XML documents.

Similarly to Section~\ref{sec:rel}, we now define the concept of 
``significantly named;'' we do it, however, for reference attributes,
rather than foreign keys.
Consider an attribute $A$ that is defined as IDREF. A DTD does not
impose any restriction on the type $E$ of an element that can be
referenced by the value of $A$. In a given XML document, $A$
(i.e.,~its name) and $E$ could be the same (e.g.,~\texttt{teacher}). If so, we
say that $A$ is an \emph{insignificantly named} reference attribute.
In the constructed data graph, the reference described by $A$ can be
represented by an implicit connector. If the opposite holds, namely,
$A$ and $E$ are different, then we say that $A$ is a
\emph{significantly named} reference attribute. In this case, the
constructed data graph should retain $A$ as the type of an explicit connector.

If attribute $A$ is defined as IDREFS, then it is insignificantly named if all
the IDs (in the value of $A$) are to elements of a type that has the same
name as $A$; otherwise, it is significantly named.

Whether a reference attribute is significantly named depends on
the given XML document (and not just on the DTD). It may change after
some future updates. As a general rule, we propose the
following. It is safe to assume that a reference attribute $A$
is significantly named if there is no element of the DTD, such that its type
is the same as $A$.  In any other case, it is best to get some human confirmation
before deciding that a reference attribute is insignificantly named.

Let $E_1$ and $E_2$ be element types.
We say that $E_2$ is a \emph{child element type} of $E_1$ if the DTD has a rule
for $E_1$ with $E_2$ on its right side.  In this case, $E_1$ is a
\emph{parent element type} of $E_2$.

Rudimentary rules for transforming an XML document to a data graph
were given in~\cite{icdtS13}.  However, they are applicable only to
simple cases.  Next, we describe a complete transformation that
consists of two stages.  We assume that prior to these two stages,
both the DTD and the XML document are examined to determine
for each reference attribute whether it is significantly named or not.

In the first stage, we analyze the DTD and classify element types as either
objects, connectors or properties. This also induces a classification
over the elements themselves.  That is, when a type $E$ is classified as
an object, then so is every element of type $E$ (and similarly when
$E$ is classified as a connector or a property).  In the second stage,
the classification is used to construct the data graph from the given
XML document.
The first stage starts by classifying all the
element types $E$ that satisfy one of the following \emph{base rules}.
\begin{enumerate}
\item\label{BR1}
If $E$ does not have any child element type
and all of its attributes are plain, then $E$ is a property.
\item\label{BR2}
If $E$ has an ID attribute or a significantly named reference attribute,
then it is an object.
\item\label{BR3}
If $E$ has neither any child element type nor an ID attribute, 
but it does have some reference attributes and all of them 
are insignificantly named, then $E$ is a connector. 
\end{enumerate}

As an example, consider the DTD of Figure~\ref{fig:dtd_example}.  Base
Rule~\ref{BR2} implies that the element types \texttt{country},
\texttt{province}, \texttt{river} and \texttt{confluence} are objects,
because the first three have an ID attribute and the fourth has a
significantly named IDREFS attribute (i.e.,~\texttt{rivers}). No base
rule applies to \texttt{economy}.  By Base Rule~\ref{BR1}, all the
other element types are properties.

Next, we find all the element types that should be classified as
properties by applying the following recursive rule. If (according to
the DTD rules) element type $E$ only has plain attributes and all of its
child element types are already classified as properties, then so is $E$.
It is easy to show that a repeated application of this recursive rule
terminates with a unique result. 

Continuing with the above example, a single application of the recursive rule 
shows that \texttt{economy} is a property, because all of its child elements
have already been classified as such by Base Rule~\ref{BR1}.

Now, we apply the following generalization of Base Rule~\ref{BR3}.  If
$E$ does not have an ID attribute, all of its child element types are
classified as properties, and it has some reference attributes and all
of them are insignificantly named, then $E$ is a connector.

We end the first stage by classifying all the remaining element types as
objects, and then
the following observations hold.
First, if an element type is classified as a property, then so
are all of its descendants.  Second, the classification (when combined
with the construction of the data graph that is described below)
ensures that a connector is always between two objects. Third, if an
element type is classified as a connector, then it has some reference
attributes and all of them are insignificantly named.

\input{fig_dtd_xml_example}

In the second stage, we transform the XML document to a data graph.
At first, we handle PCDATA as follows.  If an element $e$ (of the
document) includes PCDATA as well as either sub-elements or
attributes, then we should create a new attribute having an
appropriate name (e.g.,~\texttt{text}) and make the PCDATA its value.
This is not needed if $e$ has neither sub-elements nor attributes,
because in this case, $e$ becomes (in the data graph constructed below)
a non-nested property, such that the element type of $e$ is the name of that
property and the PCDATA is its value.

Now we construct the data graph as follows.
For each element $e$, such that $e$ is not classified as a
property, we generate a node $n_e$. This node is either an object or a
connector (and hence an explicit one) according to the classification
of $e$.  The type of $n_e$ is the same as that of $e$. If $n_e$ is an
object, we should choose one of its properties (which will be
created by the rules below) as its name.  As usual, we prefer a property
(e.g.,~\texttt{title}) that describes the meaning of $n_e$, even if it is not
a unique identifier. For each $n_e$, we create properties and
add additional edges and nodes by applying the
following six \emph{construction rules}.

\begin{enumerate} 
\item\label{pro1}
  Every plain attribute of $e$ is a property of $n_e$.
\item\label{pro2} For each child $p$ of $e$, such that $p$ is
  classified as a property, the subtree (of the given document) that
  is rooted at $p$ becomes a property of $n_e$. Note that this
  property is nested if $p$ has either plain attributes or descendants
  of its own. Also observe that element types and attribute names
  appearing in $p$ become names of properties nested in $n_e$.
\item\label{hier-edge} For each child $o$ of $e$, such that $o$ is classified as an
  object (hence, so is $e$), we add an edge from $n_e$ to $n_o$ (which is the node
  created for $o$). 
\item\label{ref-edge1} For each child $c$ of $e$, such that $c$ is classified as a
  connector, we add an edge from $n_e$ to $n_c$.
  Observe that if such a $c$ exists, then $e$ is
  classified as an object and $n_c$ is the node of the explicit connector
  corresponding to $c$.
\item\label{ref-edge2} For each reference attribute $R$ of $e$, we create new
  connectors or add edges to existing ones, according to the
  following two cases.  First, if $R$ is insignificantly named, then for each
  object $o$ that (the value of) $R$ refers to, we add an edge from
  $n_e$ to $o$.  Note that this edge is an implicit connector if $n_e$
  is an object; otherwise, it is part of the explicit connector $n_e$.

  The second case applies when $R$ is significantly named.  In this
  case, the classification rules imply that $n_e$ is an object.  We
  first create a node $n_r$, such that its only incoming edge is from
  $n_e$. This node represents an explicit connector that gets the name
  of attribute $R$ as its type and has no properties. In addition, for
  each object $o$ that (the value of) $R$ refers to, we add an edge
  from $n_r$ to $o$.
\end{enumerate}

Figure~\ref{fig:rdb_datagraph} without the dash-dotted (but with the dotted
and solid) parts shows the data graph created from the XML document
of Figure~\ref{fig:xml_example} with the DTD of
Figure~\ref{fig:dtd_example}. The two differences from the relational
transformation of Section~\ref{sec:rel} are the following.
First, the implicit connector between \texttt{France} and \texttt{Rh\^one Alpes}
is from the former to the latter (because the latter is a child of the former).
Second, there is only one explicit connector \texttt{rivers} instead of 
\texttt{river1} and \texttt{river2}.

We divide the original edges (i.e.,~those created by the above
transformation) into two kinds.  The \emph{hierarchical edges} are
those created by Construction Rule~\ref{hier-edge}. They are implicit
connectors that reflect the parent-child relationship between XML
elements.  The \emph{reference edges} are the ones introduced by
Construction Rule~\ref{ref-edge2} (i.e.,~due to reference attributes).
Construction Rule~\ref{ref-edge1} creates edges due to the element
hierarchy, but they enter nodes of explicit connectors; hence, we also
refer to them as reference edges.

As in the relational case (see Section~\ref{sec:rel}), we add opposite
edges. However, our experience indicates that even if it is done just
for the reference edges (i.e.,~no opposite edges are added for the
hierarchical ones), we generally do not miss meaningful answers to
queries.  Furthermore, as we show in the next section, a strategy that
works well 
is to assign higher
weights to opposite edges than to original ones. In this way, relevant
answers are likely to be generated first without having too many
duplicates early on.

%% file: fig_dtd_xml_example.tex
\begin{figure}[thb]
\centering
\scalebox{.8}{
\begin{minipage}{0.6\textwidth}
	\begin{tikzpicture}[->,>=stealth',shorten >=0pt,auto,node distance=10pt]
	\tikzstyle{node}=[rectangle,draw,minimum size=20pt]
	\tikzstyle{connector}=[rectangle,rounded corners, draw,minimum size=20pt]
		\setlength{\baselineskip}{9pt}
		\node[node, align=left]{
			\stl\xmltype{<!ELEMENT} \xmlattr{country} \texttt{(name,population,}\\
			\stl\ltab\texttt{economy,province)}\xmltype{>} \\
			\stl\xmltype{<!ATTLIST} \xmlattr{country} \texttt{(code ID \xmlval{\#REQUIRED}} \\		
			\stl\ltab\texttt{area CDATA \xmlval{\#IMPLIED})}\xmltype{>} \\
	
			\stl\xmltype{<!ELEMENT} \xmlattr{economy} \texttt{(gdp,inflation)}\xmltype{>}\\
	
			\stl\xmltype{<!ELEMENT} \xmlattr{province} \texttt{(name,area)}\xmltype{>} \\
			\stl\xmltype{<!ATTLIST} \xmlattr{province} \texttt{(id ID \xmlval{\#REQUIRED})}\xmltype{>} \\
			
			\stl\xmltype{<!ELEMENT} \xmlattr{river} \texttt{(name,length)}\xmltype{>}\\
			\stl\xmltype{<!ATTLIST} \xmlattr{river} \texttt{(id ID \xmlval{\#REQUIRED}}\xmltype{>} \\	
			
			\stl\xmltype{<!ELEMENT} \xmlattr{confluence} \texttt{(lng,lat)}\xmltype{>}\\
			\stl\xmltype{<!ATTLIST} \xmlattr{confluence} \\
			\stl\ltab\texttt{rivers IDREFS \xmlval{\#REQUIRED})} \\
			\stl\ltab\texttt{province IDREF \xmlval{\#REQUIRED})}\xmltype{>} \\
			
			\stl\dtdelem{name} \\
			\stl\dtdelem{population} \\
			\stl\dtdelem{gdp} \\
			\stl\dtdelem{inflation} \\
			\stl\dtdelem{area} \\
			\stl\dtdelem{length} \\
			\stl\dtdelem{lng} \\
			\stl\dtdelem{lat}
		};
			  
	\end{tikzpicture}
	\caption{\label{fig:dtd_example}DTD snippet of Mondial}
\end{minipage}
\hfill
\begin{minipage}{0.45\textwidth}
	\begin{tikzpicture}[->,>=stealth',shorten >=0pt,auto,node distance=10pt]
		\tikzstyle{node}=[rectangle,draw,minimum size=20pt]
		\setlength{\baselineskip}{9pt}
		\node[node, align=left]{
			\stl\xmltype{<country} \xmlav{code}{F} \xmlav{area}{547030}\xmltype{>} \\
			\stl\tab\xmlelem{name}{France} \\
			\stl\tab\xmlelem{population}{58M} \\
			\stl\tab\xmltype{<economy>}\\
			\stl\tab\tab\xmlelem{gdp}{\$37,728} \\
			\stl\tab\tab\xmlelem{inflation}{1.7\%} \\
			\stl\tab\xmltype{</economy>}\\
					
			\stl\tab\xmltype{<province} \xmlav{id}{prov-France-25}\xmltype{>}\\	
			\stl\tab\tab\xmlelem{name}{Rh\^one Alpes} \\
			\stl\tab\tab\xmlelem{area}{43698} \\
			\stl\tab\xmltype{</province>} \\
			\stl\xmltype{</country>}\\		
			
			\stl\xmltype{<river} \xmlav{id}{riv-Saone}\xmltype{>}\\	
			\stl\tab\xmlelem{name}{Sa\^one} \\
			\stl\tab\xmlelem{length}{473} \\
			\stl\xmltype{</river>} \\
	
			\stl\xmltype{<river} \xmlav{id}{riv-Rhone}\xmltype{>}\\	
			\stl\tab\xmlelem{name}{Rh\^one} \\
			\stl\tab\xmlelem{length}{813} \\
			\stl\xmltype{</river>} \\
	
			\stl\xmltype{<confluence} \\
			\stl\tab\tab\xmlav{province}{prov-France-25} \\
			\stl\tab\tab\xmlav{rivers}{riv-Saone riv-Rhone}\xmltype{>}\\	
			\stl\tab\xmlelem{lng}{45$^\circ$43'N} \\
			\stl\tab\xmlelem{lat}{4$^\circ$49'E} \\
			\stl\xmltype{</confluence>}			
		};
			  
	\end{tikzpicture}
	\caption{\label{fig:xml_example}XML snippet of Mondial}
\end{minipage}
}
\end{figure}

%% file: comparing.tex
\section{A Comparison}\label{sec:comparing}
In this section, we compare the data graphs produced from relational and XML
data sources. At first, we describe the example about students, courses
and lecturers that will be used.

\input{fig_compare2.tex}
\input{fig_compare_xml.tex}

We abbreviate words by their first letter as follows: S(tudent),
C(ourse), L(ecturer), E(nrolled) and G(rade).  The three entity types
student, course and lecturer have relations denoted by \texttt{S},
\texttt{C} and \texttt{L}, respectively.  The attributes of those
relations are not important.  We only assume that each entity has a
key and a name.  By a slight abuse of notation, for each of these
three relations, we will use its name also for denoting its
key. Hence, the relationship between students, courses and lecturers
is described by the relation
$\texttt{E}(\underline{\texttt{S}},\underline{\texttt{C}},
\underline{\texttt{L}},\texttt{G})$, where the attributes of the key
are underlined. The data graph constructed according to
Section~\ref{sec:rel} is given in Figure~\ref{fig:scl_rel1} (assuming that
each relation has a single tuple). Opposite edges are shown as dashed arrows.

The relation
$\texttt{E}(\underline{\texttt{S}},\underline{\texttt{C}},
\underline{\texttt{L}},\texttt{G})$ involves three entity types,
because a course may have several lecturers and not all of them teach
every student attending the course. If courses are divided into sections,
such that each one has its own lecturer(s), we can decompose
$\texttt{E}$ into two binary relationships.  To incorporate sections,
we will use the following abbreviations: A(ttend), T(each) and
Sec(tion); that is, section is abbreviated by its first three letters.
Now, the relation
$\texttt{E}(\underline{\texttt{S}},\underline{\texttt{C}},
\underline{\texttt{L}},\texttt{G})$ is replaced with
$\texttt{A}(\underline{\texttt{S}},\underline{\texttt{C}},
\underline{\texttt{Sec}},\texttt{G})$ and
$\texttt{T}(\underline{\texttt{C}},
\underline{\texttt{Sec}},\underline{\texttt{L}})$.  Note that a section
is a weak entity, and each of the new relations has two foreign
keys, where one of them consists of two attributes (i.e.,~$\texttt{C}$
and $\texttt{Sec}$) that together uniquely identify a section of a
course; that is, the value of $\texttt{Sec}$ is just a number, such as
$1$, $2$, etc.  The data graph for the five relations $\texttt{S}$,
$\texttt{C}$, $\texttt{L}$, $\texttt{A}$ and $\texttt{T}$ is given in
Figure~\ref{fig:scl_rel2}.

The data graph produced from XML is shown in Figure~\ref{fig:scl_xml}
(the XML document and its DTD are not shown due to a lack of space).
Figure~\ref{fig:scl_xml} has more data (e.g.,~two lecturers) than 
Figures~\ref{fig:scl_rel1} and~\ref{fig:scl_rel2} to illustrate
some points later.  It has opposite edges (depicted as dashed
arrows) only for reference (but not hierarchical) edges.  One
rectangle (for Section 1) and two edges are dotted. They are additions
to the data graph that will be explained later.  For now, they should
be ignored (hence, there is an edge from DB directly to each 
\texttt{enrolled} connector).

To show the differences between the three data graphs, we consider the
query $\{\texttt{Student}, \texttt{Lecturer}\}$.  On the data graph of
Figure~\ref{fig:scl_rel1}, one answer has only original edges, an
$\texttt{enrolled}$ node as the root, and $\texttt{Smith}$ and
$\texttt{Ullman}$ as the leaves. This answer is likely to be generated
early on for two reasons. First, it is as small as can be (i.e.,~only
three nodes). Second, it benefits from a strategy of assigning higher
weights to opposite edges than original ones. In more detail, some algorithms
(e.g.,~\cite{icdeBHNCS02,vldbKPCSDK05,icdtGS16}) enumerate answers
in an order that is correlated with increasing weight. If
the most relevant answers are likely to have only original edges,
then those algorithms would find them early on 
when opposite edges have higher weights.

For each course in which $\texttt{Smith}$ is taught by
$\texttt{Ullman}$, we would get an answer with an $\texttt{enrolled}$
node as the root, and those two as the leaves. Either we remove
duplicates by treating $\texttt{enrolled}$ nodes according to their
type, rather than id (i.e.,~all of them are identical to one another), or we should add
the course so that users can grasp how seemingly duplicate answers are
different from one another.  In this case, we need to augment each
answer with only one additional node, namely, the \texttt{course}
object (pointed to by the $\texttt{enrolled}$ connector). This
requires adding only one node to each answer. The main drawback of the
data graph of Figure~\ref{fig:scl_rel1} is a large number of
$\texttt{enrolled}$ connectors, since they represent a ternary
relationship.

On the data graph of Figure~\ref{fig:scl_rel2}, the answer with
$\texttt{Smith}$ and $\texttt{Ullman}$ as the leaves must use a
mixture of original and opposite edges, and has five nodes. We need to
add a sixth node if we want to show how duplicates are different from
one another. In comparison with Figure~\ref{fig:scl_rel1}, answers are
larger implying that it takes longer to find them. Moreover, the
strategy of assigning higher weights to opposite edges than original
ones is not effective, because the most relevant answers have both
types (so this approach would not help in generating them early on).

Next, we consider the data graph of Figure~\ref{fig:scl_xml}, which is
obtained from XML. At first, we ignore the dotted rectangle and two
edges.  The answer that $\texttt{Ullman}$ teaches $\texttt{Smith}$
consists of only original edges and three nodes; a fourth one is needed
to show the course. This is the same as in
Figure~\ref{fig:scl_rel1}, but the number of $\texttt{enrolled}$
connectors is smaller (i.e.,~equal to the number of $\texttt{attend}$
connectors in Figure~\ref{fig:scl_rel2}, where $\texttt{teach}$ connectors 
are also used). The main advantage
of Figure~\ref{fig:scl_xml}, however, is its flexibility.
If a lecturer teaches all the students enrolled in the course (which is
likely to be true in many cases), then it is sufficient to have a
connector from $\texttt{course}$ to $\texttt{lecturer}$, such as the
dotted edge from $\texttt{DB}$ to $\texttt{Vardi}$ (i.e.,~no need for
edges between that lecturer and the $\texttt{enrolled}$ connectors of
students attending the course).  Now,
the subtree $\texttt{Jones}
\leftarrow \texttt{enrolled} \leftarrow \texttt{DB} \rightarrow
\texttt{Vardi}$ is the answer that $\texttt{Vardi}$ teaches $\texttt{Jones}$.
It consists of four nodes and already shows the
course, which means that duplicates cannot occur.  If we introduce
sections (e.g.,~the dotted rectangle) and add edges to their lecturers
(e.g.,~the dotted arrow pointing to $\texttt{Ullman}$), the subtree
$\texttt{Ullman} \leftarrow 1 \rightarrow \texttt{enrolled} \rightarrow
\texttt{Smith}$ is the answer that $\texttt{Ullman}$ teaches $\texttt{Smith}$.
It consists of four nodes, and a fifth one should be added to show the course.

To summarize, a data graph obtained from XML has the following
advantages over one constructed from a relational database.
\begin{enumerate}
\item Answers have an equal or smaller number of nodes when the same
  information (e.g.,~sections) is represented in both cases.
\item Relevant answers are more likely to use only original edges.
\item The data graph requires fewer nodes to represent ternary
  relationships (e.g.,~enrolled), because of the XML hierarchy.
\item The biggest advantage is heterogeneity:
\begin{itemize}
\item It is sufficient to have sections only in courses that have
  more than one of them.
\item We can directly link a lecturer to a course, section or
  individual students depending on how she is assigned.
\item Thereby, we reduce the size of the data graph, give rise to fewer
  duplicates, and make answers more meaningful, because they show how
  the lecturer is assigned.
\end{itemize}
\end{enumerate}

%% file: fig_compare2.tex
\tikzstyle{node}=[rectangle,draw,minimum size=20pt]
\tikzstyle{connector}=[rectangle,rounded corners, draw,minimum size=20pt]

\begin{figure}[thb]
\centering
\begin{minipage}{0.35\textwidth}

\begin{tikzpicture}[->,>=stealth',shorten >=0pt,auto,node distance=.5cm]  
  \node[connector, align=center] (e) at (0,0) {enrolled\\ grade : $90$};
  \node[node] (s) [below left=.3 and -.6 of e] {Smith {\scriptsize (student)}};
  \node[node] (l) [below right=.3 and -.6 of e] {Ullman {\scriptsize (lecturer)}};
  \node[node] (c) [above = of e] {DB {\scriptsize (course)}};
  
  \draw (e) edge [bend right = 15] (s);
  \draw (s) edge [dashed, bend right = 15] (e);
  \draw (e) edge [bend right = 15] (l);
  \draw (l) edge [dashed, bend right = 15] (e);
  \draw (e) edge [bend right = 15] (c);
  \draw (c) edge [dashed, bend right = 15] (e);
\end{tikzpicture}
\caption{\label{fig:scl_rel1}Data graph from RDB with one ternary relationship}
\end{minipage}
\hfill
\begin{minipage}{0.60\textwidth}
\begin{tikzpicture}[->,>=stealth',shorten >=0pt,auto,node distance=1cm]  
  \node[node] (sec) at (0,0) {1 {\scriptsize (section)}};
  \node[node] (s) [left = .5 of sec] {Smith {\scriptsize (student)}};
  \node[node] (l) [right = .5 of sec] {Ullman {\scriptsize (lecturer)}};

  \node[connector, align=center] (a) [above = 0.9 of s] {attend\\ grade : $90$};
  \node[node] (c) [above = of sec] {DB {\scriptsize (course)}};
  \node[connector] (t) [above = of l] {teach};

  \draw (a) edge [bend right = 15] (s);
  \draw (s) edge [dashed, bend right = 15] (a);
  \draw (a) edge [bend left = 15] (sec);
  \draw (sec) edge [dashed, bend left = 15] (a);
  \draw (c) edge [bend right = 15] (sec);
  \draw (sec) edge [dashed, bend right = 15] (c);
  \draw (t) edge [bend right = 15] (sec);
  \draw (sec) edge [dashed, bend right = 15] (t);
  \draw (t) edge [bend right = 15] (l);
  \draw (l) edge [dashed, bend right = 15] (t);
\end{tikzpicture}
\caption{\label{fig:scl_rel2}Data graph from RDB with binary relationships}
\end{minipage}
\end{figure}

%% file: fig_compare_xml.tex
\tikzstyle{node}=[rectangle,draw,minimum size=20pt]
\tikzstyle{connector}=[rectangle,rounded corners, draw,minimum size=20pt]

\begin{figure}[t]
\centering
\begin{tikzpicture}[->,>=stealth',shorten >=0pt,auto,node distance=0.3cm]  
  \node[node] (c) at (0,0) {DB {\scriptsize (course)}};
  \node[node] (ll) [left = 1 of c] {Ullman {\scriptsize (lecturer)}};
  \node[node] (lr) [right = 1 of c] {Vardi {\scriptsize (lecturer)}};
  \node[node, densely dotted] (sec) [below = of c] {1 {\scriptsize (section)}};
  \node[connector, align=center] (e80) [below left = of sec] {enllored \\ grade : $80$};
  \node[connector, align=center] (e90) [below right = of sec] {enllored \\ grade : $90$};
  \node[node] (s80) [left = 1 of e80] {Smith {\scriptsize (student)}};
  \node[node] (s90) [right = 1 of e90] {Jones {\scriptsize (student)}};  
  
  \draw (ll) edge [dashed, bend left = 15] (e80);
  \draw (e80) edge [bend left = 15] (ll);
  \draw (e80) edge [bend right = 15] (s80);
  \draw (s80) edge [dashed, bend right = 15] (e80);
  
  \draw (lr) edge [dashed, bend left = 15] (e90);
  \draw (e90) edge [bend left = 15] (lr);
  \draw (e90) edge [bend right = 15] (s90);
  \draw (s90) edge [dashed, bend right = 15] (e90);
  
  \draw (c) edge (sec);
  \draw (sec) edge (e80);
  \draw (sec) edge (e90);

  \draw  (c) edge [densely dotted] (lr);
  
  \draw (sec)[densely dotted] edge (ll);
\end{tikzpicture}
\caption{\label{fig:scl_xml}Data graph from XML}
\end{figure}
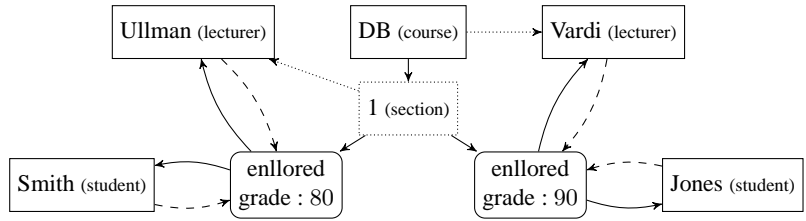

%% file: conclusion.tex
\section{Conclusions}
We showed that the OCP model is an effective conceptual basis for
constructing data graphs. Using it, we developed transformations for generating
data graphs from RDB and XML. These transformations are quite elaborate and
provide much better results than the ad hoc methods that have been used in
the literature thus far. In particular, the produced data graphs are better
in terms of both efficiency (i.e.,~answers are generated more quickly)
and effectiveness (i.e.,~the most relevant answers are produced early on).

It should be emphasized that the presented transformations are based
on the principle of creating fat nodes (as explained in
Section~\ref{sec:advantages}) and avoiding redundancies (e.g.,~due to
insignificantly named references).  Thus, they are applicable and
useful (in most if not) all cases, regardless of how answers are
generated or ranked.

We showed that XML is the preferred starting point for constructing
data graphs. However, we need to better understand how to create XML
documents that yield the best possible data graphs. Toward this end,
we plan to develop appropriate design rules for XML documents.

Due to space limitations, we did not discuss how to generated data
graphs from RDF. In our experience, it is harder to do that than when
starting with RDB or XML. An important principle of RDF is unique
representation by means of URIs (uniform resource identifiers). As a
result, RDF triples are highly fragmented (e.g.,~there could be a
separate triple for storing each person's title, such as Dr., Mrs.,
etc.), which makes it hard to create a coherent data graph with fat nodes.

An interesting topic for future work is 
to how to construct data graphs from XML documents without DTDs.

%% file: main.bbl
\begin{thebibliography}{10}
\providecommand{\url}[1]{\texttt{#1}}
\providecommand{\urlprefix}{URL }

\bibitem{sigmodAGKS10}
Achiezra, H., Golenberg, K., Kimelfeld, B., Sagiv, Y.: Exploratory keyword
  search on data graphs. In: SIGMOD Conference (2010)

\bibitem{dblpBLCL09}
Bao, Z., Ling, T.W., Chen, B., Lu, J.: Effective {XML} keyword search with
  relevance oriented ranking. In: ICDE (2009)

\bibitem{icdeBHNCS02}
Bhalotia, G., Hulgeri, A., Nakhe, C., Chakrabarti, S., Sudarshan, S.: Keyword
  searching and browsing in databases using {BANKS}. In: ICDE (2002)

\bibitem{CW14}
Coffman, J., Weaver, A.C.: An empirical performance evaluation of relational
  keyword search techniques. {IEEE} Trans. Knowl. Data Eng.  26(1),  30--42
  (2014)

\bibitem{pvldbDKS08}
Dalvi, B.B., Kshirsagar, M., Sudarshan, S.: Keyword search on external memory
  data graphs. PVLDB  (2008)

\bibitem{icdeDYWQZL07}
Ding, B., Yu, J.X., Wang, S., Qin, L., Zhang, X., Lin, X.: Finding top-k
  min-cost connected trees in databases. In: ICDE (2007)

\bibitem{sigmodGKS08}
Golenberg, K., Kimelfeld, B., Sagiv, Y.: Keyword proximity search in complex
  data graphs. In: SIGMOD Conference (2008)

\bibitem{pvldbGKS11}
Golenberg, K., Kimelfeld, B., Sagiv, Y.: Optimizing and parallelizing ranked
  enumeration. PVLDB  (2011)

\bibitem{dexawGS16}
Golenberg, K., Sagiv, Y.: Constructing data graphs for keyword search. In: DEXA
  (2016), presentation url:
  {https://drive.google.com/open?id=0BxX7DI4NO\_-vTFZaRGgzU25WdjQ}

\bibitem{icdtGS16}
Golenberg, K., Sagiv, Y.: A practically efficient algorithm for generating
  answers to keyword search over data graphs. In: ICDT (2016)

\bibitem{sigmodGSBS03}
Guo, L., Shao, F., Botev, C., Shanmugasundaram, J.: {XRANK}: Ranked keyword
  search over {XML} documents. In: SIGMOD Conference (2003)

\bibitem{sigmodHWYY07}
He, H., Wang, H., Yang, J., Yu, P.S.: {BLINKS}: ranked keyword searches on
  graphs. In: SIGMOD Conference (2007)

\bibitem{icdeHPB03}
Hristidis, V., Papakonstantinou, Y., Balmin, A.: Keyword proximity search on
  {XML} graphs. In: ICDE (2003)

\bibitem{vldbKPCSDK05}
Kacholia, V., Pandit, S., Chakrabarti, S., Sudarshan, S., Desai, R.,
  Karambelkar, H.: Bidirectional expansion for keyword search on graph
  databases. In: VLDB (2005)

\bibitem{icdeKRSSW09}
Kasneci, G., Ramanath, M., Sozio, M., Suchanek, F.M., Weikum, G.: {STAR}:
  Steiner-tree approximation in relationship graphs. In: ICDE (2009)

\bibitem{sigmodLOFWZ08}
Li, G., Ooi, B.C., Feng, J., Wang, J., Zhou, L.: {EASE}: an effective 3-in-1
  keyword search method for unstructured, semi-structured and structured data.
  In: SIGMOD Conference (2008)

\bibitem{edbtMS16}
Mass, Y., Sagiv, Y.: Virtual documents and answer priors in keyword search over
  data graphs. In: Proceedings of the Workshops of the {EDBT/ICDT} 2016 Joint
  Conference (2016)

\bibitem{ipmPL15}
Park, C., Lim, S.: Efficient processing of keyword queries over graph databases
  for finding effective answers. Inf. Process. Manage.  51(1),  42--57 (2015)

\bibitem{icdtS13}
Sagiv, Y.: A personal perspective on keyword search over data graphs. In: ICDT.
  pp. 21--32 (2013)

\bibitem{edbtXP08}
Xu, Y., Papakonstantinou, Y.: Efficient {LCA} based keyword search in {XML}
  data. In: EDBT (2008)

\end{thebibliography}
